\documentclass{aa}
\usepackage{graphics}
\input{psfig}
\begin{document}

\thesaurus{02(02.14.1), 06(06.05.1, 06.09.1, 06.15.1)}

\title{Solar Models and NACRE thermonuclear reaction rates}

\author{P. Morel\inst{1}, B. Pichon\inst{2}, J. Provost\inst{1}
 \and G. Berthomieu\inst{1}}

\institute{
D\'epartement Cassini, UMR CNRS 6529, Observatoire de la C\^ote 
d'Azur, BP 4229, 06304 Nice CEDEX 4, France \and
DARC, UMR CNRS 8629, Observatoire de Paris Section Meudon, 92195 Meudon
CEDEX, France.
}

\offprints{P. Morel}
\mail{Pierre.Morel@obs-nice.fr}

\date{Received date / Accepted date}

\maketitle

\begin{abstract}
Using the most recent updated physics,
calibrated solar models have been computed with the new
thermonuclear reaction rates of NACRE,
the recently available European compilation.
Comparisons with models computed with the reaction
rates of Caughlan \& Fowler (\cite{cf88}) and
of Adelberger et al. (\cite{a98}) are made for
global structure, expected neutrinos fluxes, chemical composition and
sound speed profiles, helioseismological properties
of p-modes and g-modes.
 
\keywords{Physical data and processes: nuclear reactions,
nucleosynthesis, abundances --
Sun: evolution -- Sun: interior -- Sun: Oscillations }
\end{abstract}

\section{Introduction}\label{sec:int}
Precise solar models have been constructed over the past three and a half
decades (see, e.g. Bahcall et al. \cite{b63}; Bahcall \cite{b93},
Bahcall et al. \cite{bbp98}).
The refinement have accelerated in the past
decade (see, e.g. Bahcall and Ulrich \cite{bu88}, Bahcall \& Pinsonneault
\cite{bbp92}).
Since ten years many stellar and solar
models have been computed using the
thermonuclear reaction rates of
Caughlan \& Fowler (\cite{cf88}, hereafter C88)
a popular compilation but not optimized for solar conditions.
More recently some authors (e.g. Bahcall et al.
\cite{bp95}; Reiter et al. \cite{rww95}; Chaboyer et al. \cite{cdp95};
Berthomieu et al. \cite{bpm95}; Christensen-Dalsgaard et al. \cite{c96})
employed the improved
thermonuclear reaction rates adopted by Bahcall \& Pinsonneault (\cite{bp92})
for the calculation of accurate solar models. 
Meanwhile several groups of nuclear
physicists have undergone other compilations of updated thermonuclear
reaction rates of astrophysical interest. 
A year ago the compilation of
Adelberger et al. (\cite{a98}, hereafter A98) has been published.
The original motivation of this compilation is to assess the state of
the nuclear physics important to the solar neutrino problem.
The incidences on solar models of these new reaction rates have
been analyzed by several groups (see e.g. Bahcall et al. \cite{bbp98}; Brun et al. \cite{btm98};
Morel et al. \cite{mpb98}). More recently
the European Nuclear Astrophysics Compilation of REaction rates
(Angulo et al. \cite{a99}, NACRE, hereafter N99)
has been completed and opened to free access.
The driving motivation of this last
work, coordinated by the Institut d'Astronomie et d'Astrophysique of the
Universit\'e Libre de Bruxelles, is the build-up of well documented
and evaluated sets of experimental data or theoretical
predictions of astrophysical interest. 
To have a direct idea of the degree of reliability 
of any reaction rate, the authors have published, either
a very convenient plot of the available
measurements of cross section S-factors with respect to energy,
or a table showing the range of the various
parameters needed for cross section evaluation, e.g. the resonance parameters.
Moreover, the accuracy of each analytical fit is indicated.
This new compilation gives besides the adopted reaction rate $\cal R$, its
lower and  upper limits $\cal R_{\rm l}$ and $\cal R_{\rm u}$.
A solar neutrino analysis based on preliminary NACRE data for the PP reactions
has been done by Castellani et al. (\cite{c97}). Recently Arnould et al.
(\cite{agj99}) have used the N99 reaction rates to compute abundance
predictions in non hydrogen and helium burning. They convincingly
show that large spreads in the
abundances predictions for several nuclides may result not only from a change
in temperature, but also from nuclear physics uncertainties.

\begin{table*}
\caption[]{
Respectively for the C88, A98 and N99 compilations, the 
S-factors at zero energy S(0) (MeV barn) and S'(0) (barn) are given for
each nuclear reaction of solar interest, 
but $\element[][7]{Be}(e^-,\nu_{\element[][7]{Be}}\gamma)\element[][7]{Li}$.
The last column gives $\Delta\cal R$, the global uncertainty of N99's rates
computed, for $T_6=15$, according to Eq.~\ref{eq:prec}.
The S(0) factor of $\element[][12]{C}(p,\gamma)\element[][13]{N}$ is estimated
from the N99 plot.
}\label{tab:0}
\begin{tabular}{llllllllllllll}  \hline \\
&\multicolumn{2}{c}{C88}&
\multicolumn{2}{c}{A98}&
\multicolumn{3}{c}{N99}\\
reactions   & S(0)  & S'(0)  & S(0)&S'(0) & S(0)&S'(0)&$\Delta{\cal R}$\\
\\ \hline \\
$\element[][1]{H}(p,\beta^+\nu_{\rm pp})\element[][2]{H}$&
$4.06\,10^{-25}$&$4.6\,10^{-24}$&
$4.00\,10^{-25}$&$4.48\,10^{-24}$&
$3.94\,10^{-25}$&$4.61\,10^{-24}$&5\% \\ \\
$\element[][2]{H}(p,\gamma)\element[][3]{He}$&
$2.5\,10^{-7}$&$7.9\,10^{-6}$&&&
$2.0\,10^{-7}$&$5.6\,10^{-6}$&40\%\\ \\
$\element[][3]{He}(\element[][3]{He},2p)\element[][4]{He}$&
$5.56$&$-8.2$&
$5.4$&$-4.1$&
$5.18$&$-2.22$&6\%\\ \\
$\element[][3]{He}(\alpha,\gamma)\element[][7]{Be}$&
&&$5.3\,10^{-4}$&$-3.0\,10^{-4}$&
$5.4\,10^{-4}$&$-5.2\,10^{-4}$&18\%\\ \\
$\element[][7]{Li}(p,\alpha)\element[][4]{He}$&
&&&&$5.93\,10^{-2}$&$0.193$&14\%\\ \\
$\element[][7]{Be}(p,\gamma)\element[][8]{B}^*$&
$2.39\,10^{-5}$&$0.$&
$1.9\,10^{-5}$&$-1.35\,10^{-5}$&
$2.1\,10^{-5}$&$-1.8\,10^{-5}$&11\%\\ \\
$\element[][12]{C}(p,\gamma)\element[][13]{N}$&
$1.40\,10^{-3}$&$4.2\,10^{-3}$&
$1.34\,10^{-3}$&$2.6\,10^{-3}$ &$1.5\,10^{-3}$&&11\%\\ \\
$\element[][13]{C}(p,\gamma)\element[][14]{N}$&
$5.5\,10^{-3}$&$1.3\,10^{-2}$&
$7.6\,10^{-3}$&$-7.8\,10^{-3}$ &
$7.0\,10^{-3}$&&23\%\\ \\
$\element[][14]{N}(p,\gamma)\element[][15]{O}$&
$3.2\,10^{-3}$&$-5.7\,10^{-3}$&
$3.5\,10^{-3}$&$-1.28\,10^{-2}$ &$3.2\,10^{-3}$&&34\%\\ \\
$\element[][15]{N}(p,\gamma)\element[][16]{O}$&
$6.4\,10^{-2}$&$3.2\,10^{-2}$&
$6.4\,10^{-2}$&$2.1\,10^{-2}$&
$6.4\,10^{-2}$&&23\%\\ \\
$\element[][15]{N}(p,\alpha)\element[][12]{C}$&
$71.$&$423.$&
$67.5$&$310.$ &$69.$&&15\%\\ \\
$\element[][16]{O}(p,\gamma)\element[][17]{F}$&
$9.3\,10^{-3}$&&
$9.4\,10^{-3}$&$-2.4\,10^{-2}$ &
$9.3\,10^{-3}$&&36\%\\ \\
$\element[][17]{O}(p,\alpha)\element[][14]{N}$&
&&$9.58\,10^{-4}$&$1.08\,10^{-2}$&&&35\%\\ \\
\hline 
\end{tabular}
\end{table*}
We are now in the fortunate position of having two precise and {\em
independent} determinations of the best nuclear fusion data, namely A98 and
N99. In order to illustrate the effects, on the standard solar model,
of nuclear fusion rates on various
astronomical quantities, including neutrino fluxes and helioseismology
frequencies, we compare the model results calculated
with the best current data from A98
and N99, with results obtained using early estimates of fusion rates of C88. 
Those differences are not very large. Nevertheless
they modify the energy balance, the stratification, the chemical composition
and the neutrino generation in the core.

Let us first recall
the main constraints known nowadays on solar models.
The helioseismological
constraints relevant to the core are the small
p-mode frequency differences $\delta\nu_{02}$ and $\delta\nu_{13}$ and
the not yet observed spectrum of gravity modes. 
Other signatures of changes
of thermonuclear reaction rates will be the sound velocity profile
which is known from inversion of helioseismic data between
$R\ga 0.1\,R_\odot$ and $R\la 0.9\,R_\odot$, and also the radius of
the base of the solar convection zone which is precisely located.
The amount of observed photospheric depletions of lithium and beryllium
which are often ascribed to
transport phenomena beneath the convection zone
are also sensitive to changes
of the thermonuclear reaction rates in their low energy regime.
An another constraint also connected to nuclear reaction rates is
the isotopic ratio \element[][3]{He}~/~\element[][4]{He} measured
at present day at the solar surface which is sensitive
to the pre-main sequence deuteron burning and to initial isotopic ratios
\element[][2]{H}~/~\element[][1]{H} and \element[][3]{He}~/~\element[][4]{He}
of cosmological interest. 

The paper is organized as follows: in
Sec.~\ref{sec:aa} in the low energy range, for the nuclear reactions
of interest for solar modeling, we summarize the main differences between
N99 and A98 with respect to C88.
The physics used in the models is described Sec.~\ref{sec:phy}.
In Sec.~\ref{sec:comp} we report results of comparisons 
between calibrated solar models computed with N99, A98 and C88,
finally we conclude in Sec.~\ref{sec:dis}.

\begin{figure*}
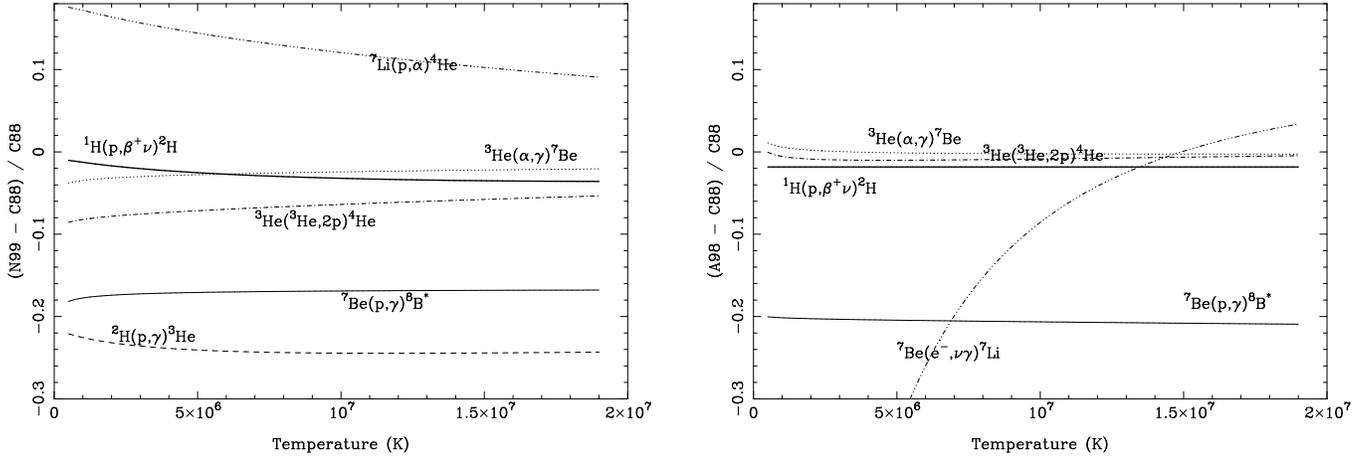

\hbox{\psfig{figure=diff1.ps,height=6cm,angle=270}
\hspace{0.5truecm}
\psfig{figure=diff3.ps,height=6cm,angle=270}}
\caption{
Left panel: relative differences between the reaction rates
between N99 and C88
for the relevant solar temperature range $0.5\leq T_6\leq 19$.
for the PP reactions:
$\element[][1]{H}(p,\beta^+\nu_{\rm pp})\element[][2]{H}$ (heavy, full),
$\element[][2]{H}(p,\gamma)\element[][3]{He}$ (dashed),
$\element[][3]{He}(\element[][3]{He},2p)\element[][4]{He}$ (dot-dash-dot-dash),
$\element[][3]{He}(\alpha,\gamma)\element[][7]{Be}$ (dotted),
$\element[][7]{Li}(p,\alpha)\element[][4]{He}$(dash-dot-dot-dot) and
$\element[][7]{Be}(p,\gamma)\element[][8]{B^*}$ (thin, full).
Right panel: same between A98 and C88.
}\label{fig:diff1}
\end{figure*}

\section{Comparison of thermonuclear reaction rates from N99, A98 and C88 
compilations}\label{sec:aa}
The  most important nuclear reactions relevant of solar modeling are,
for the PP chains (Clayton \cite{c68}; Bahcall \cite{bb89} Table 3.1 and 3.3): 
\begin{description}
 \item $\element[][1]{H}(p,\beta^+\nu_{\rm pp})\element[][2]{H}$,
 $\element[][2]{H}(p,\gamma)\element[][3]{He}$,
 $\element[][3]{He}(\element[][3]{He},2p)\element[][4]{He}$,
 \item $\element[][3]{He}(\alpha,\gamma)\element[][7]{Be}$,
 $\element[][7]{Be}(e^-,\nu_{\element[][7]{Be}}\gamma)\element[][7]{Li}$,
 $\element[][7]{Li}(p,\alpha)\element[][4]{He}$, 
 \item 
$\element[][7]{Be}(p,\gamma)\element[][8]{B}^*(\beta^+\nu_{\element[][8]{B}})\element[][8]{Be}
 (\alpha)\element[][4]{He}$,
\end{description}
 and for the CNO bi-cycle:
\begin{description} 
 \item
 $\element[][12]{C}(p,\gamma)\element[][13]{N}(\beta^+\nu_{\element[][13]{N}})
 \element[][13]{C}$,
 $\element[][13]{C}(p,\gamma)\element[][14]{N}$,
 \item
 $\element[][14]{N}(p,\gamma)\element[][15]{O}(\beta^+\nu_{\element[][15]{O}})
 \element[][15]{N}$,$\element[][15]{N}(p,\gamma)\element[][16]{O}$,
 $\element[][15]{N}(p,\alpha)\element[][12]{C}$,
 \item
 $\element[][16]{O}(p,\gamma)\element[][17]{F}(\beta^+\nu_{\element[][17]{F}})
 \element[][17]{O}$,
 $\element[][17]{O}(p,\alpha)\element[][14]{N}$.
\end{description}
Owing to their low termination and small
contribution to energetic and nucleosynthesis,
despite their interest for neutrino generation, we do not explicitly
take into account in the nuclear network
$\element[][1]{H}(p e^-,\nu_{\rm pep})\element[][2]{H}$
and $\element[][3]{He}(p, e^+\nu_{\rm hep})\element[][4]{He}$
the so-called $pep$ and $hep$ reactions. Nevertheless
we compute the number of $\nu_{\rm pep}$ neutrino
generated using the equation (3.17) of the Bahcall's
(\cite{bb89}) reference text book.

\begin{figure*}
\hbox{\psfig{figure=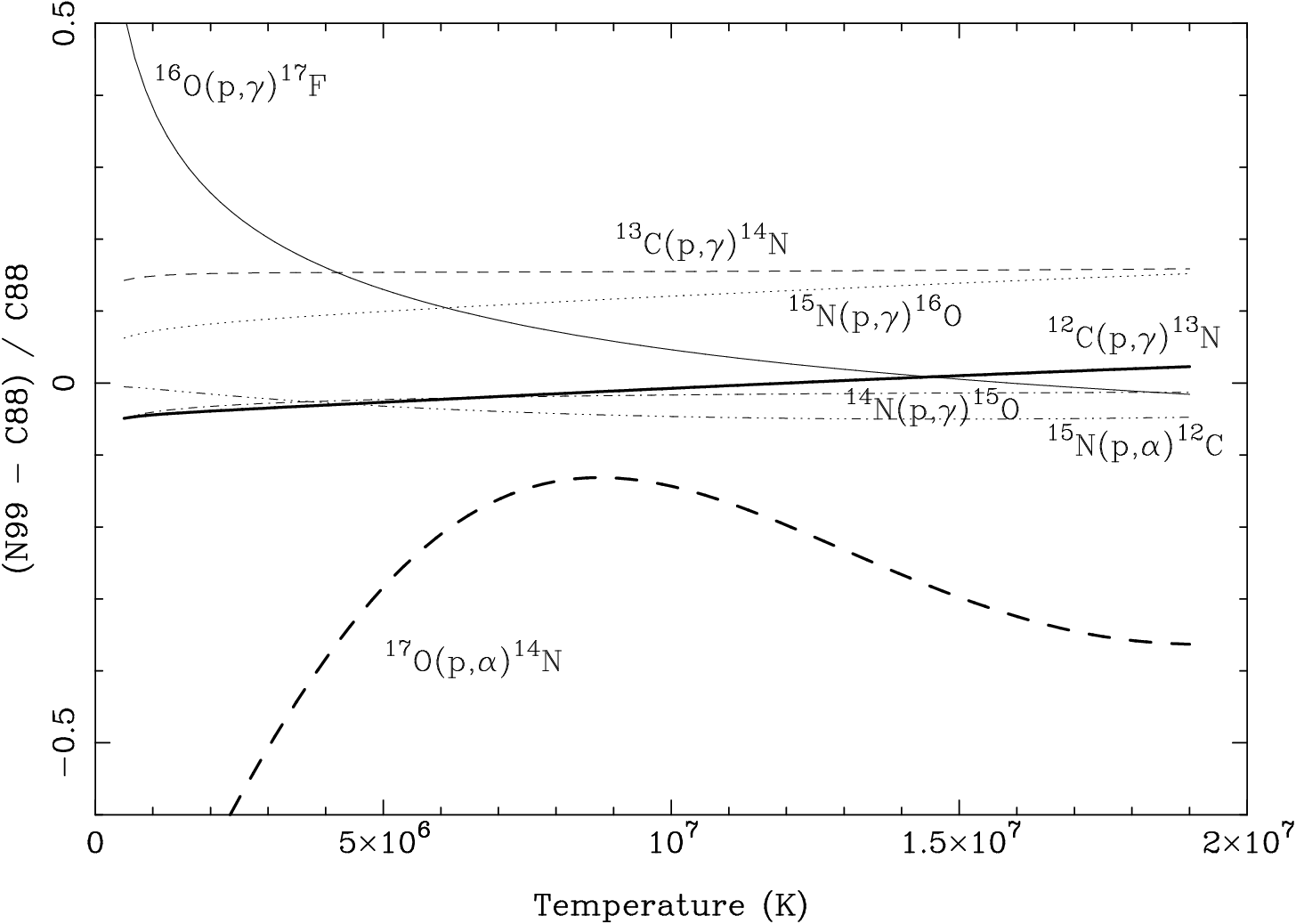,height=6cm,angle=270}
\hspace{0.5truecm}
\psfig{figure=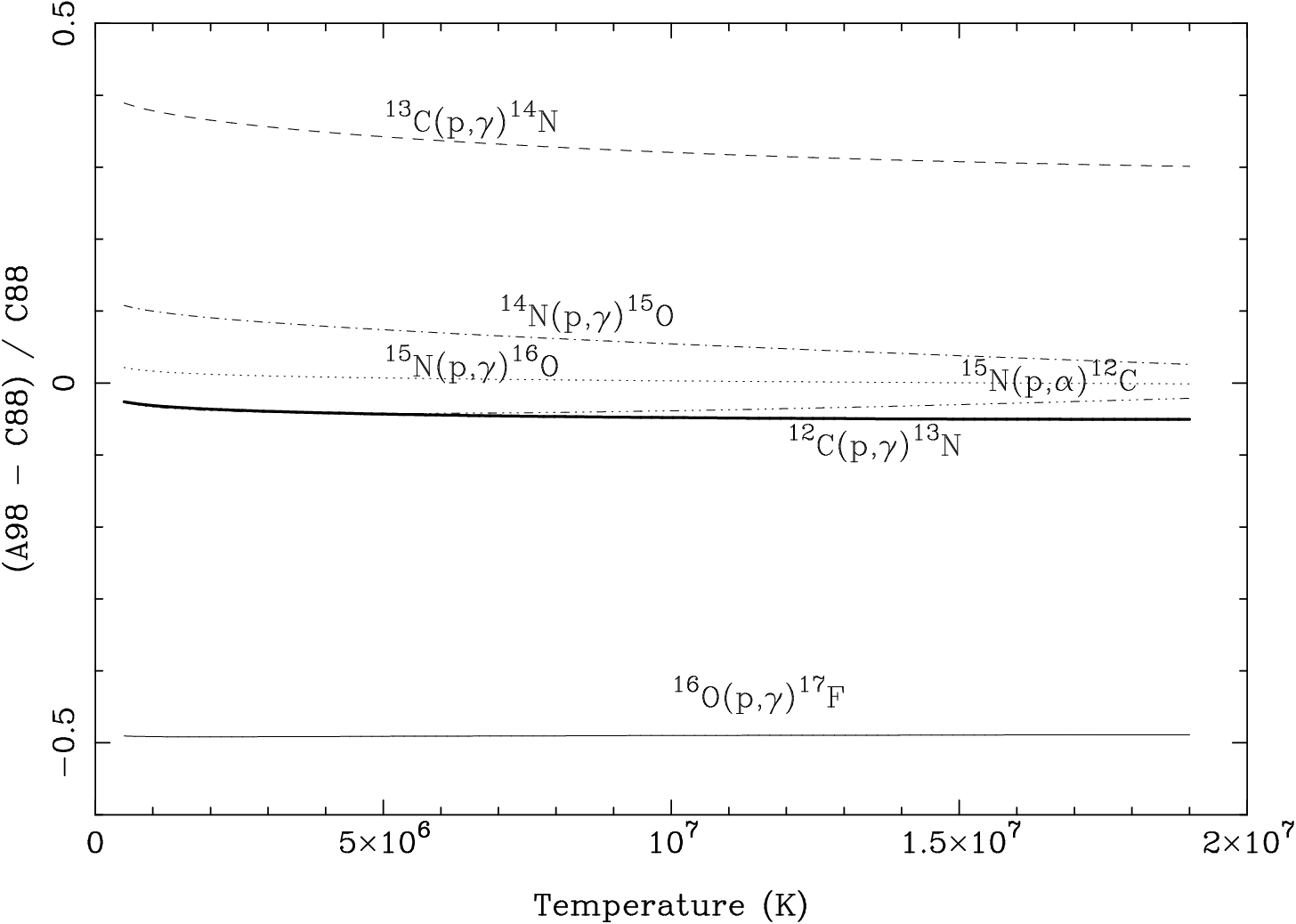,height=6cm,angle=270}}
\caption{
The same as Fig.~\ref{fig:diff1} for the CNO reactions:
$\element[][12]{C}(p,\gamma)\element[][13]{N}$ (heavy, full),
$\element[][13]{C}(p,\gamma)\element[][14]{N}$ (dashed),
$\element[][14]{N}(p,\gamma)\element[][15]{O}$ (dot-dash-dot-dash),
$\element[][15]{N}(p,\gamma)\element[][16]{O}$ (dotted),
$\element[][15]{N}(p,\alpha)\element[][12]{C}$ (dash-dot-dot-dot),
$\element[][16]{O}(p,\gamma)\element[][17]{F}$ (thin, full),
$\element[][17]{O}(p,\alpha)\element[][14]{N}$ (heavy, dashed).
}\label{fig:diff2}
\end{figure*}

The changes between the reaction rates of N99, A98 and C88 are extensively
commented in Adelberger et al. (\cite{a98}) and Angulo et al. (\cite{a99}).
As a matter of illustrations, for the three compilations
 and for each PP and CNO reaction -- but the electronic capture
 $\element[][7]{Be}(e^-,\nu_{\element[][7]{Be}}\gamma)\element[][7]{Li}$ --
Table~\ref{tab:0} gives the S-factors at zero energy and
the underlying global
uncertainty on the rate $\Delta{\cal R}$:
\begin{equation}\label{eq:prec}
\Delta{\cal R} = \sqrt{\frac{\cal R_{\rm u}}{\cal R_{\rm l}}} - 1,
\end{equation}
estimated for $T_6=15$; $T_6$ is the temperature in M\,K, 
$\cal R_{\rm l}$ ($resp.$ $\cal R_{\rm u}$) stands for 
lower ($resp.$ upper) limit of N99 updated reactions.
For our thermonuclear
reaction network the contributions of resonances to the astrophysical reaction
rates are negligible in the solar range of temperatures,  
therefore the values of
S(0) and, if any, S'(0) presented here are pertinent. For sake of
briefly we omit to reproduce the known S"(0) values.
Figure~\ref{fig:diff1}
($resp.$ Fig.~\ref{fig:diff2}) compares the relative differences
between the adopted rates of N99 ($resp.$ A98) and C88
for the temperature range $0.5 \leq T_6\leq 19$. 

We next briefly recall the main changes in the rates of A98 and N99,
with respect to those of C88 which is the oldest and, up to nowadays,
 the most used and complete.
 
\underline{$\element[][2]{H}(p,\gamma)\element[][3]{He}$}:
Among all reactions of PP chains and CNO bi-cycle, it is the rate of this
PPI reaction which is the most badly known.
The reaction rate is so fast that it is only involved by the
pre-main sequence deuteron burning.
Owing to the lower value adopted for the S-factors at zero energy,
the rate of the
reaction which synthesizes \element[][3]{He} is about $-24\%$
lower in N99 than in C88.
This reaction is not updated in A98, for the calculations with A98
we used the value adopted in C88.

\underline{$\element[][3]{He}(\element[][3]{He},2p)\element[][4]{He}$}:
For the most energetic reaction of the PP chains
N99 ($resp.$ A98) adopts values smaller by about $-6\%$ 
($resp.$ $-2\%$) than C88 for the S-factors. As consequences of the
calibration process, for the models using either N99 or A98,
more \element[][1]{H} nuclear fuel will be burnt in order to reach, at
present day, the observed luminosity and effective temperature. Therefore
these models will have cores with larger temperature, helium content, density
and sound velocity than models computed with C88; then, at first sight, their predicted
total neutrino fluxes are expected to be larger.
This effect will be enhanced for the models computed with N99 since
the rates of the two
reactions $\element[][1]{H}(p,\beta^+\nu_{\rm pp})\element[][2]{H}$
and $\element[][2]{H}(p,\gamma)\element[][3]{He}$
are smaller in N99 than in C88.

Figure p. 26 of Angulo et al. (\cite{a99}) 
gives the impression that, at low energy,
the Junker at al.'s (\cite{j98}) recent measurements 
are avoided by the interpolation formulae adopted
(see also Fig.~2 of Adelberger et al. \cite{a98}).
These recent data should lead to an increase of
S(0) so, to an enhancement of the efficiency
of the reaction and then, owing to the calibration process, to
a decrease of solar neutrino fluxes. 
   
\underline{$\element[][7]{Li}(p,\alpha)\element[][4]{He}$}:
The S(0) value adopted by N99
differs from C88 between $+15\%$ and $+7\%$. In the core
its curate rate is irrelevant due, first to its strong rate
($\approx 10^{-5}$\,year, e.g. Bahcall
\cite{bb89} Table 3.2) and, second to the tiny mass fraction 
of $\element[][7]{Li}\sim2\,10^{-15}$.
Beneath the convection zone the burning of \element[][7]{Li}
will be more efficient with N99 than with C88, leading to an increase
of the lithium depletion at the solar surface at present day.
As suggested in C88, the \element[][7]{Li} burning is slightly
enhanced by few percents by the neighbor reaction 
$\element[][7]{Li}(p,\gamma)\element[][8]{Be}(\alpha)\element[][4]{He}$
which has been added to our nuclear network.
This reaction is not updated in A98, for the calculations with A98
we shall use the value adopted in C88.
   
\underline{$\element[][7]{Be}(e^-,\nu_{\element[][7]{Be}}\gamma)\element[][7]{Li}$}:
N99 deals only with charged particle induced reactions
involving nuclei, and therefore 
the \element[][7]{Be} electron capture rate is not updated.
In the calculations with N99
we shall used the value given by A98.
Beneath $T_6=1$ only an upper limit is given
in C88 for the \element[][7]{Be} electron capture.
 The adopted rate of A98 differs
from the rate of C88 by more than $+50\%$ at low temperature;
 for $T_6\sim 15$, i.e. in the solar core, the rates of C88 and A98
are of same order.

\underline{$\element[][7]{Be}(p,\gamma)\element[][8]{B}^*$}:
This reaction controls the efficiency of the important source
of $\nu_{\element[][8]{B}}$, the so-called boron solar neutrino. 
The adopted values for the S-factors at zero energy are slightly larger
in N99 than in A98, but still smaller than in C88.   
With respect to C88, everything else equal, one can expect
that the neutrino flux from boron will be
{\em reduced} for the solar models computed with A98 and N99.
 
\begin{figure*}
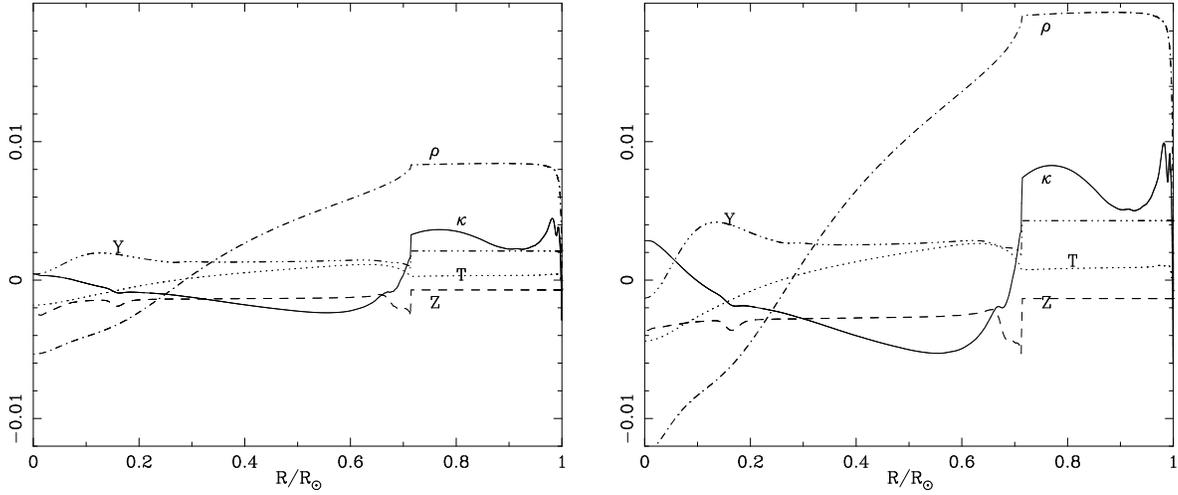

\hbox{\psfig{figure=diff_A.ps,height=6.5cm,angle=270}
\hspace{0.5truecm}
\psfig{figure=diff_C.ps,height=6.5cm,angle=270}}
\caption{
Relative differences in opacity $\kappa$ (full), heavy element $Z$ (dashed),
density $\rho$ (dot-dash-dot-dash), temperature $T$ (dotted) and helium content
$Y$ (dash-dot-dot-dash)
for the calibrated models A98 (left) and C88 (right), with respect to model N99.
}\label{fig:AC}
\end{figure*}
   
\underline{$\element[][13]{C}(p,\gamma)\element[][14]{N}$}:
The values of S-factors at zero energy adopted by N99 and A98 are
magnified by a factor of about two with respect to their previous values in C88;
as a result the rates are increased by $+30\%$ and $+15\%$ respectively. These
large differences will not have any noticeable incidence 
on the global structure of the core since the energy generated
by the CNO bi-cycle is only $\la2\%$ of the total nuclear energy.

\underline{$\element[][14]{N}(p,\gamma)\element[][15]{O}$}:
The rate of the most important reaction
for the computation of energy generation and neutrino fluxes
created by the CNO bi-cycle is known with a large uncertainty. The three
compilations adopt about the same values for the S-factors at zero energy.
Figure~\ref{fig:diff2} shows
small differences between the rates. This is due to 
different interpolation formulas which slightly differ since
there is no measurement at low energy (see the convincing
figure p. 58 of Angulo et al. (\cite{a99})).

\underline{$\element[][15]{N}(p,\gamma)\element[][16]{O}$}:
For the reaction which governs the efficiency of the
NO-part of the CNO bi-cycle, N99, A98 and  C88 adopt
the use of S-factors at zero energy
obtained by Rolfs \& Rodney (\cite{rr74}). Due to differences
in the interpolation formula
Fig.~\ref{fig:diff2} reveals 
enhanced rates of $+15\%$ in N99 with respect to C88 or A98.


\underline{$\element[][16]{O}(p,\gamma)\element[][17]{F}$}:
At low energy the reaction which controls the generation
of $\nu_{\element[][17]{F}}$, the so-called fluorine solar neutrino, is based
on data with large experimental errors. The adopted rate has the largest
uncertainty among the CNO reactions. Though Table~\ref{tab:0} gives for
 the three compilations about
the same values for the S-factors, Fig.~\ref{fig:diff2}
shows large differences for the rates resulting from different
analytical formulations. Beyond $T_6~\sim 10$, N99 and C88 are
close (Angulo et al. \cite{a99}).
The difference of $-50\%$ between A98 and C88 results of
the used of the
standard formulation of the non-resonant reaction rate with S-factors
(Fowler et al. \cite{fgz67}).

\underline{$\element[][17]{O}(p,\alpha)\element[][14]{N}$}:
N99 and A98 use different analytical fits based on the measurements of
Landr\'e et al. (\cite{l89}). They differ by $-30\%$.
With the discovery of a resonance at low energy (Landr\'e et al. $loc.\,cit.$)
the analytical fit of C88 became in error by more than
two order of magnitude. For the models computed with C88
we have used the rates derived from the Landr\'e's et al. analytical fit,
as recommended by A98.

\paragraph{Summary.} With respect to C88 many reaction rates, principally
$\element[][3]{He}(\element[][3]{He},2p)\element[][4]{He}$, are lowered
in N99 and also, but in a less extend, in A98. One can expect that this
will lead to
calibrated solar models with central cores with {\em larger} temperature.
For the reactions of PP chains, with respect to C88, other important
changes connected to the observable neutrino fluxes are the rates of
the electronic capture $\element[][7]{Be}(e^-,\nu_{\element[][7]{Be}}\gamma)\element[][7]{Li}$
which is significantly diminished for $T_6 \la7$ in A98 and, for N99,
the decrease of the rate of $\element[][7]{Be}(p,\gamma)\element[][8]{B}^*$.
With respect to C88, the changes in N99 and A98 of 
the reaction rates of the CNO bi-cycle
are not large enough to modify significantly the solar model.

\begin{figure*}
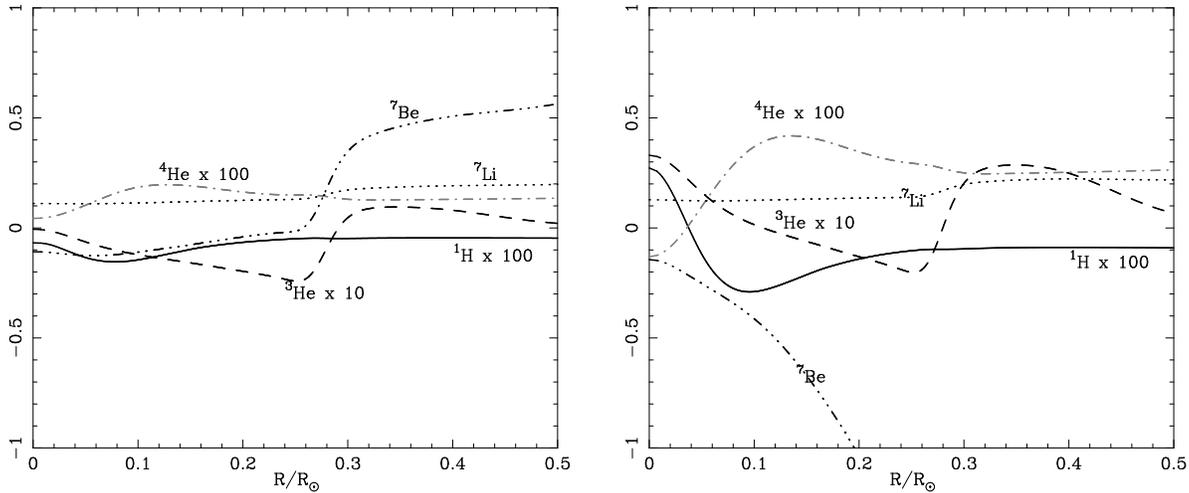

\hbox{\psfig{figure=diff1A.ps,height=6.5cm,angle=270}
\hspace{0.5truecm}
\psfig{figure=diff1C.ps,height=6.5cm,angle=270}}
\caption{
Relative differences of abundances with respect to model N99,
for models A98 (left) and C88 (right) for PP species:
$\element[][1]{H}\times 100$ (full),
$\element[][3]{He}\times 10$ (dashed),
$\element[][4]{He}\times 100$ (dash-dot-dash),
$\element[][7]{Li}$ (dotted),
\element[][7]{Be} (dash-dot-dot-dash).
}\label{fig:dpp}
\end{figure*}

\begin{figure*}
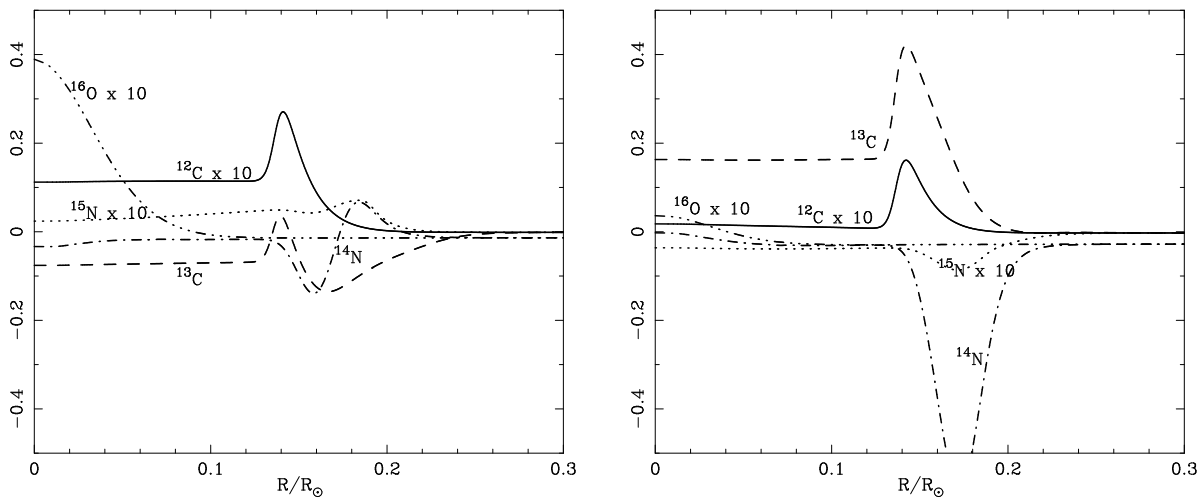

\hbox{\psfig{figure=diff2A.ps,height=6.5cm,angle=270}
\hspace{0.5truecm}
\psfig{figure=diff2C.ps,height=6.5cm,angle=270}}
\caption{
Same as Fig.~\ref{fig:dpp} for the CNO species:
$\element[][12]{C}\times 10$ (full),
$\element[][13]{C}$ (dashed), 
$\element[][14]{N}$ (dash-dot-dash-dot),
$\element[][15]{N}\times 10$ (dotted),
$\element[][16]{O}\times 10$ (dash-dot-dot-dash).
}\label{fig:dcno}
\end{figure*}

\section{The solar models}\label{sec:phy}
Basically the physics of the models is the same as in
Morel et al. (\cite{mpb97}).
\paragraph{Calibration of models.}
Each evolution is initialized with a homogeneous zero-age
pre-main-sequence model in quasi-static gravitational contraction
with the temperature at center $T_{\rm c}\sim0.5$\,MK,
i.e. close to the onset of the deuteron burning.
The models are calibrated within
a relative accuracy better than $10^{-4}$ by adjusting:
the ratio $l/H_{\rm p}$ of the mixing-length to the 
pressure scale height, the initial mass fraction $X_{\rm i}$
of hydrogen and the initial mass fraction $(Z/X)_{\rm i}$ of
heavy element to hydrogen in order that, at present day,
the solar models have the luminosity $L_\odot=3.846\,10^{33}$\,erg\,s$^{-1}$
(Guenther et al. \cite{gdkp92}), the radius
$R_\odot=6.9599\,10^{10}$\,cm  (Guenther et al. $loc.\,cit.$)
and the mass fraction of heavy element to hydrogen $(Z/X)_\odot=0.0245$
(Grevesse \& Noels \cite{gn93}). We used a time of evolution 
$t_{\rm ev}=4600$\,My, an intermediate value between
the meteoritic age  $t_{\odot\rm m}=4530\pm 40$\,My
of the Sun\footnote{Here $t_{\odot\rm m}$ is referenced with respect
to ZAMS which occurs $36\pm10$\,My (Guenter $loc.\,cit.$)
after the formation of meteorites
$4566\pm5$\,My from now (Bahcall et al. \cite{bp95}).}
(Guenther \cite{g89}) 
and its helioseismic value $t_{\odot\rm h}=4660\pm100$\,My derived by
Dziembowski et al. (\cite{dfrs98}).
The zero age main-sequence (ZAMS) is defined as the time where nuclear
reactions dominate gravitation as the primary energy source
by more than 50\% (Guenther et al. $loc.\,cit.$).
The mass of the Sun is assumed to be
$M_\odot=1.9891\,10^{33}$\,g  (Cohen \& Taylor \cite{ct86}).

\paragraph{Nuclear and diffusion network.}
The general nuclear network we used contains the following species~:
\element[][1]{H},
\element[][2]{H},
\element[][3]{He},
\element[][4]{He},
\element[][7]{Li},
\element[][7]{Be},
\element[][9]{Be},
\element[][12]{C},
\element[][13]{C},
\element[][14]{N},
\element[][15]{N},
\element[][16]{O},
\element[][17]{O} and 
\element[][]{Ex}; \element[][]{Ex} is an ``Extra'' fictitious
mean non-CNO heavy element with atomic mass 28 and charge 13
($\element[][]{Ex}\sim\element[][28]{Al}$) which
complements the mixture i.e.,
$X_{\element[][]{Ex}}=1-\sum_{i=\element[][1]{H}}^{\element[][17]{O}}X_i$
with $X_i$ as the mass fraction of the species labeled with
$i=\element[][1]{H},\ldots,\element[][17]{O}$.
With respect to time, due to microscopic diffusion processes, the
abundances of heavy elements are enhanced toward the center;  
\element[][]{Ex}  mimics that enhancement for the non CNO metals 
which contribute to changes of $Z$, then to opacity variations but
neither to nuclear energy generation nor to nucleosynthesis.
To compute the depletion of \element[][9]{Be},
we have added, to the nuclear network given Sec.~\ref{sec:aa},
the most efficient reactions of \element[][9]{Be} burning:
$\element[][9]{Be}(p,d)2\,\element[][4]{He}$ and 
$\element[][9]{Be}(\alpha,n)\element[][12]{C}$.
The life time of the neutron, namely 888\,s (Barnett et al. \cite {b96}),
is smaller by more than thirteen orders of magnitude than
the evolutionary time scale of the
Sun's main-sequence $pp$ reaction. Therefore, for the calculations,
the last reaction is rewritten
$\element[][9]{Be}(\alpha,e^-p\bar\nu_{\element[][9]{Be}})\element[][12]{C}$.
The weak screening of Salpeter (\cite{s54}) is used,
it is a very good approximate of the exact solution
of the Schr\"odinger equation for the fundamental
$pp$ reaction (Bahcall et al. \cite{bck98}).

The protosolar initial isotopic ratios (in number) for hydrogen and helium
are respectively taken as
$\element[][2]{H}/\element[][1]{H}=3.01\,10^{-5}$,	
\element[][3]{He}/\element[][4]{He}$=1.1\,10^{-4}$
(Gautier \& Morel \cite{gm97}). 
The initial ratios between the heavy elements within $Z$ are set to
their photospheric present day values, namely  
(in number) C: 0.24551, N: 0.06458 and O: 0.51295
(Grevesse \& Noels \cite{gn93}) then, for the complement Ex: 0.17696.
The initial isotopic ratios are derived from
the abundances of nuclides (Anders \& Grevesse \cite{ag89}):
\element[][13]{C}/\element[][12]{C}$=1.11\,10^{-3}$,
\element[][15]{N}/\element[][14]{N}$=4.25\,10^{-3}$,
\element[][17]{O}/\element[][16]{O}$=3.81\,10^{-4}$.
We have used the meteoritic values (Grevesse \& Sauval \cite{gs98}) for the
initial abundances in dex, ($\element[][]{H}\equiv 12$),
of \element[][]{Li} and \element[][]{Be}:
\[ \left[\frac{\element[][]{Li}}{\element[][]{H}}\right]=3.31\pm0.04,
\ \left[\frac{\element[][]{Be}}{\element[][]{H}}\right]=1.42\pm0.04.\]
For the calculations of depletions, the lithium
is assumed to be in its most abundant isotope \element[][7]{Li} form,
so it is with beryllium assumed to be \element[][9]{Be}.
Neither the meteoritic abundance nor the nuclide isotopic ratio of
\element[][7]{Be} are known,
 due to numerical constraints 
the protosolar abundance of 
\element[][7]{Be} was somehow arbitrarily taken to a very low,
but non zero value, namely
$[\element[][7]{Be}/\element[][1]{H}]=-3.58$\,dex.
The initial abundance of each isotope is derived from
isotopic ratios and initial values of
$X\equiv \element[][1]{H}+\element[][2]{H}$,
$Y\equiv \element[][3]{He}+\element[][4]{He}$ and $Z/X$ as inferred
by the calibration process in order to fulfill 
the basic relationship $X+Y+Z\equiv1$.

Microscopic diffusion is described by the simplified formalism of 
Michaud \& Proffitt (\cite{mp93}) with each of the heavy elements as a
trace element.

\paragraph{Equation of state, opacities, convection and atmosphere.}
We have used the OPAL equation of state (Rogers et al.
\cite{rsi96}) and opacities (Iglesias \& Rogers \cite{ir96})
for the solar mixture of Grevesse \& Noels (\cite{gn93}) 
complemented, at low temperatures, respectively by the MHD equation of state
(D\"appen \cite{d96}) and
Alexander \& Ferguson (\cite{af94}) opacities.
The interpolations of opacities are made with
the v9 birational spline package of G. Houdek
(Houdek \& Rogl \cite{hr96}; Houdek \cite{h98}).

In the convection zones the temperature gradient is
computed according to the standard mixing-length theory.
The mixing-length is defined as $l\equiv \alpha H_{\rm p}$,
where $H_{\rm p}$ is the pressure scale height.
The convection zones are mixed via
a strong turbulent diffusion  coefficient, which produces a homogeneous
composition.

The atmosphere is restored using a $T(\tau)$ law derived
from an atmosphere model of the Sun computed by van't Veer (\cite{v98})
with the Kurucz's (\cite{k91}) ATLAS12 package.
The connection with the envelope is made at the Rosseland optical
depth $\tau_{\rm b} = 20$ (Morel et al. \cite{m94}), where the diffusion approximation for radiative
transfer becomes valid. A smooth connection of the
gradients is insured between the uppermost layers of the envelope
and the optically thick convective part of the atmosphere.
The radius $R_\star$ of any model
is taken at the optical depth $\tau_\star\simeq 0.54$ where
$T(\tau_\star)=T_{\rm eff}$;
the mass of the star $M_\star$, is defined as the mass enclosed in
 the sphere of radius $R_\star$.
The external boundary is located at the optical
depth $\tau_{\rm ext}=10^{-4}$, where the density is fixed to its value in
the atmosphere model $\rho(\tau_{\rm ext})=3.55\,10^{-9}$\,g\,cm$^{-3}$,
that corresponds about to the temperature minimum in the solar chromosphere.

\paragraph{Numerics.}
The models have been computed using the CESAM code (Morel \cite{m97}).
The numerical schemes are fully implicit and their accuracy
is first order for the time and third order for the space.
For numerical performance and algorithmic constraints
the analytical expressions of reaction rates are tabulated
with respect to temperature for the range $0.5\leq T_6\leq  20$
and interpolated with a relative accuracy better than $10^{-5}$.
Each evolution needs about 90 models.
Typically 600 mass shell are used along the evolution,
it increases up to 2100 for the models used in seismological analysis.

\paragraph{p-mode and g-mode oscillation calculations.}
The frequencies of linear, adiabatic, global acoustic modes of the
solar models have been computed for degrees $\ell=0$ to $\ell=150$
and have been compared to the observations.
The characteristic low degree p-mode frequency
differences $\Delta\nu_{n,\ell}=\nu_{n,\ell}-\nu_{n-1, \ell+2}$ 
for $\ell=0$ and $\ell=1$,
which provide information on the properties of the solar core,
have been fitted by linear regressions with respect to $n$: 
\[
\Delta\nu_{n,\ell}= \delta\nu_{n,\ell} + S_\ell (n-n_0),\ n_0=21,\ \ell =0,1,
\] 
both for the observations and the theoretical frequencies. 
For the gravity modes which have not yet been observed, we give the 
characteristic asymptotic spacing period $P_0$ according to Provost 
\& Berthomieu (\cite{pb86}). 

\begin{table*}
\caption[]{ 
Comparison of global characteristics of solar models computed with the
thermonuclear reaction rates of respectively Angulo et al.
(\cite{a99}, N99),
Adelberger et al. (\cite{a98}, A98) and Caughlan \& Fowler
(\cite{cf88}, C88) compilations. The first four rows give the initial
values of $Y_{\rm i}$ ($resp.$ $Z_{\rm i}$) the mass fraction of
helium ($resp.$ heavy element),
$(Z/X)_{\rm i}$ the mass fraction of heavy element to hydrogen,
$\alpha$ the mixing-length parameter.
$\element[][7]{Li}_{\rm sz}$ ($resp.$ $\element[][9]{Be}_{\rm sz}$)
and $\element[][7]{Li}_{\rm s}$ ($resp.$ $\element[][9]{Be}_{\rm s}$)
respectively are the surface depletions
in dex ($\element[][]{H}\equiv 12$) of
\element[][7]{Li} ($resp.$ \element[][9]{Be}) 
at zero-age main sequence and at present day. All other item are for present 
day.
The surface isotopic ratio $(\element[][3]{He}/\element[][4]{He})_{\rm s}$
is in unit of $10^{-4}$; 
$Y_{\rm s}$, $Z_{\rm s}$ respectively are
the surface mass fraction of helium and heavy element;
$R_{\rm CZ}$ is the radius, in solar units, at the bottom of
the convection zone;
$T_{\rm c}$, $\rho_{\rm c}$, $Y_{\rm c}$ and $Z_{\rm c}$ are the central
values respectively of, the temperature
in units of $10^7$K, the density in g\,cm$^{-3}$, the mass fraction,
of helium and heavy element.
$\delta\nu_{02}$ and $\delta\nu_{13}$
are the values, in $\mu$Hz, of the frequency
differences between the radial p-modes of degree $\ell=0-2$ and $\ell=1-3$.
$P_0$ is the characteristic spacing period of g-modes in minutes.
}\label{tab:1}
\begin{tabular}{llllll} \\  \hline \\
                             & N99   & A98    & C88  & Observed values\\
\\ \hline \\
$Y_{\rm i}$                  &0.2723 &0.2726 &0.2729 \\
$Z_{\rm i}$                  &0.0197 &0.0197 &0.0196 \\
$(Z/X)_{\rm i}$              &0.0278 &0.0278 &0.0277 \\
$\alpha$	             &1.924  &1.931  &1.941  \\ \\
$\element[][7]{Li}_{\rm sz}$ &2.26  & 2.38 & 2.37 \\
$\element[][9]{Be}_{\rm sz}$ &1.42  &1.42  & 1.42 \\ \\
$\element[][7]{Li}_{\rm s}$  &2.18  &2.30  &2.29  &$1.10\pm0.10$, Grevesse \& 
Sauval (\cite{gs98})\\
$\element[][9]{Be}_{\rm s}$  &1.35  &1.35  &1.353  &$1.40\pm0.09$, Grevesse \& 
Sauval (\cite{gs98})\\
$(\element[][3]{He}/\element[][4]{He})_{\rm s}$&4.34&4.32&4.32&$4.4\pm0.4$, 
Bodmer et al. (\cite{b95})\\
$Y_{\rm s}$                  &0.2436 &0.2442 &0.2447 &$0.232\ -\ 0.249$, Basu
(\cite{b97b})\\
$Z_{\rm s}$                  &0.0181 &0.0181 &0.0181 \\
$R_{\rm CZ}$                 &0.7138 &0.7132 &0.7124 &$0.713\pm 0.001$, Basu \& 
Antia (\cite{ba95})\\
$T_{\rm c}$                  &1.573  &1.570  &1.566  \\
$\rho_{\rm c}$               &153.8  &153.0  &151.9  \\
$Y_{\rm c}$                  &0.6418 &0.6420 &0.6409 \\
$Z_{\rm c}$                  &0.0210 &0.0210 &0.0210 \\ \\
$\delta\nu_{02}$  	     & 9.21  & 9.18  & 9.16  &$9.002\pm 0.044\ -\ 
9.014\pm0.042$, from LOI/GOLF (see text)\\
$\delta\nu_{13}$  	     &16.10  & 16.06 & 16.03 &$15.884\pm 0.034\ -\ 
15.711\pm0.071$, from LOI/GOLF (see text)\\ 
$P_{0}$                      &35.13  & 35.23 & 35.42 \\
\\ \hline \\
\end{tabular}
\end{table*}

\section{Comparison of models}\label{sec:comp}
Table~\ref{tab:1} gives the global properties of models and Fig.~\ref{fig:AC}
exhibits the profiles, with respect to radius, of the most important variables
for the internal structure namely, density, temperature,
opacity, helium and heavy element contents.

\subsection{Chemical composition}
The changes in chemical composition directly result from changes of
thermonuclear reaction rates but also, in a more intricate way, from 
changes in microscopic diffusion coefficients which are sensitive to the
temperature and density, and to
chemical composition, pressure, temperature and density gradients.
 
\paragraph{Changes at the surface and in the envelope.} For the three models
N99, A98 and C88, Table~\ref{tab:1} shows that the 
expected photospheric abundances of helium are slightly reduced and remain
 compatible with the range
of observed values.
As known (Basu \cite{b97b}), the amount of photospheric
``observed'' helium derived from inversion of
helioseismic data is more sensitive to the equation of state
than the amount of photospheric ``predicted'' helium derived from
calibrated solar models. Indeed
we have calibrated a solar
model\footnote{Not analyzed here for sake of briefness.}
using C88 thermonuclear reaction rates and
the MHD (D\"appen \cite{d96}) equation of state instead of
OPAL and obtained a
photospheric helium content $Y_{\rm s}=0.246$ which is the value
derived from inversion using the MHD equation of state
(Basu \& Antia \cite{ba95}).

Though the \element[][7]{Li} surface depletion is increased with the use of
the enhanced rate of $\element[][7]{Li}(p,\alpha)\element[][4]{He}$
adopted by N99, the predicted
abundance is still very far from the observed value.
The differences are likely consequences of the lack, in
standard solar models, of mixing generated by the shear at the level
of the tachocline (see e.g. Gough et al.~\cite{g96}, Brun et al. \cite{btz99}).
It smoothes the chemical composition gradients and reduces the
microscopic diffusion efficiency
(Basu \cite{b97a}) immediately beneath the convection zone.
According to the new observations (Grevesse \& Sauval \cite{gs98}),
the predicted photospheric depletion of beryllium is tiny.  
The predictions for the surface isotopic ratios
$(\element[][3]{He}/\element[][4]{He})_{\rm s}$ 
by the three models are all
within the interval of accuracy given by the observations.

\paragraph{Changes in the core.}
The solar core is the innermost part where
the nuclear energy generation
is efficient. It extends from the center to about $R_{\rm c}\simeq 0.4R_\odot$
slightly beyond the \element[][3]{He} peak located around $0.3R_\odot$.
Owing to the less efficiency of PP reactions,
see Sec.~\ref{sec:aa},
the temperature, the density, the amount of helium and the sound velocity
at center of {\em calibrated} models N99 and A98, are larger than in C88. 
As expected, Fig.~\ref{fig:dpp} shows almost symmetrical profiles for
the differences for \element[][1]{H} and \element[][4]{He}. Owing to the
larger efficiency of the $pp$ reaction in C88, larger values
are obtained for the relative difference C88 minus N99 than for A98 minus N99.
The typical features for the relative differences of abundances of
\element[][3]{He} are consequences of the smaller reaction rates
of N99 with respect to A98 and C88 of
$\element[][3]{He}(\element[][3]{He},2p)\element[][4]{He}$. 
Beneath the \element[][3]{He} peak, owing to
the increase of the temperature, the amount of
\element[][3]{He} is smaller in C88 and A98 than in N99; it is the
reverse beyond the peak. 
Though the same rate prevails in N99 and A98 for the 
\element[][7]{Be} electronic capture,
there is a non zero value for relative difference between the 
\element[][7]{Be} profiles of models N99 and A98,
resulting of differences between the rates adopted for
$\element[][3]{He}(\alpha,\gamma)\element[][7]{Be}$.

Figure~\ref{fig:dcno} exhibits  large differences in the abundances of
\element[][16]{O} for A98 and C88 with respect to N99
despite the fact that the rates of the reactions of \element[][16]{O} burning
are close. In fact, these differences result from changes of rates of
 \element[][15]{N} burning which creates \element[][16]{O}.
For \element[][12]{C} and \element[][14]{N},
around $0.18 R_\odot$, effects of changes of nuclear reaction rates
are magnified by the large gradients of that species. There, Fig.~\ref{fig:AC}
reveals, on $Z$ profiles, small bumps due to the magnification by
large gradients of variations in chemical composition caused by the changes of
thermonuclear reaction rates.

\subsection{Thickness of the convection zone.}
For radius $R\ga 0.4R_\odot$, i.e. in the envelope,
Fig.~\ref{fig:AC} shows that
the opacity profiles are close within $\pm0.4\%$ for models N99 and A98. The  
thickness of the convection zone is about the same in N99 and A98, and close,
within the error bars (see Table~\ref{tab:1}), to the observed value.
It is slightly larger for model C88. That difference
is due to the increase
of the radiative temperature gradient resulting from the higher value
of the opacity. The differences of temperature between N99 and A98 being small,
the changes in opacity are mainly due to the variations of density. The
relative opacity differences amount to $\pm1\%$ between C88 and N99.

\begin{figure*}
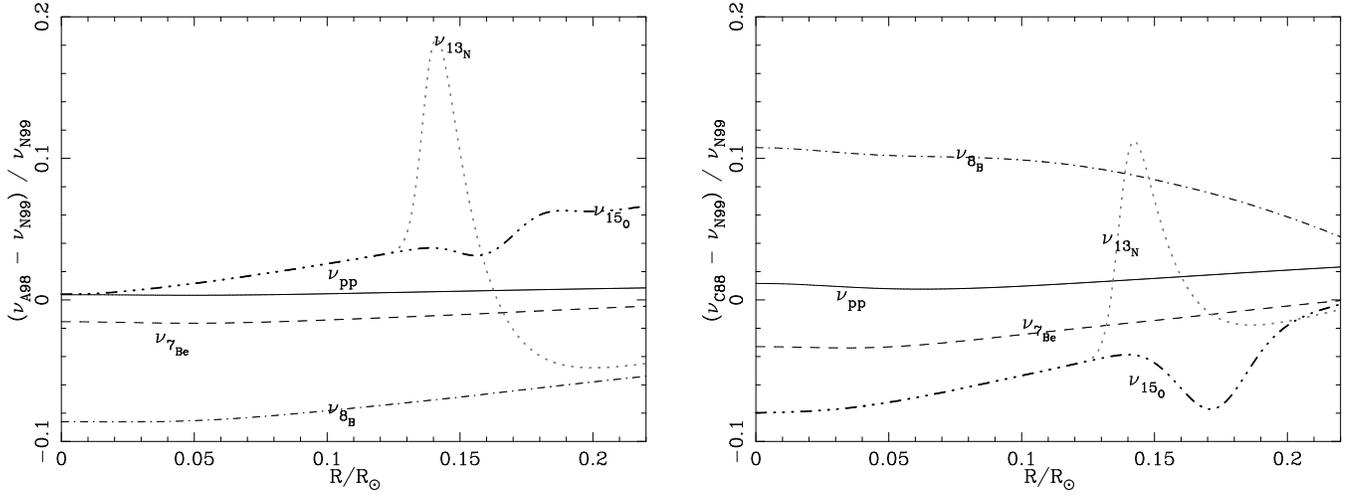

\hbox{\psfig{figure=diff_nu1.ps,height=6.5cm,angle=270}
\hspace{0.5truecm}
\psfig{figure=diff_nu2.ps,height=6.5cm,angle=270}}
\caption{Left panel: relative differences of neutrino rates generation, between
models A98 and N99, for
$\nu_{\rm pp}$ (thin, full),
$\nu_{\element[][7]{Be}}$ (dashed),
$\nu_{\element[][8]{B}}$ (dot-dash-dot-dash),
$\nu_{\element[][13]{N}}$ (dotted), 
$\nu_{\element[][15]{O}}$ (dash-dot-dot-dash).
Right panel: the same between models C88 and N99
}\label{fig:dnu}
\end{figure*}

\begin{table}
\caption[]{
The same as Table~\ref{tab:1} for the neutrino fluxes.
Using the comprehensive symbols, defined Sec.~\ref{sec:aa}, 
$\nu_{\rm pp}$, $\nu_{\rm pep}$, $\nu_{\element[][7]{Be}}$, 	
$\nu_{\element[][8]{B}}$, 
$\nu_{\element[][13]{N}}$, $\nu_{\element[][15]{O}}$,
$\nu_{\element[][17]{F}}$ 
are, at earth level, the number ${\rm cm^{-2}\ s^{-1}}$ of neutrinos
of each kind.
$\Phi_{\rm Ga}$ and $\Phi_{\rm Cl}$ in SNU and
$\Phi_{\rm Ka}$ in events day$^{-1}$, are the expected fluxes
for the three neutrino
experiments namely gallium, chlorine and Kamiokande (see text).
The observed values for $\Phi_{\rm Ga}$, $\Phi_{\rm Cl}$ and $\Phi_{\rm Ka}$
are respectively from Hampel et al. (\cite{ha99}), Davis (\cite{da94}) and Fukuda et 
al. (\cite{fu96}).
}\label{tab:2}
\begin{tabular}{llllll}  \hline \\
                             & N99   & A98    & C88   & Observed values\\
\\ \hline \\
$\nu_{\rm pp}$		     &$5.91\,10^{10}$&$5.92\,10^{10}$&$5.94\,10^{10}$\\ 
$\nu_{\rm pep}$		     &$1.40\,10^{8}$ &$1.40\,10^{8}$ &$1.48\,10^{8}$ \\ 
$\nu_{\element[][7]{Be}}$    &$4.90\,10^{9}$ &$4.80\,10^{9}$ &$4.71\,10^{9}$ \\ 
$\nu_{\element[][8]{B}}$     &$5.68\,10^{6}$ &$5.17\,10^{6}$ &$6.19\,10^{6}$ \\ 
$\nu_{\element[][13]{N}}$    &$5.73\,10^{8}$ &$5.77\,10^{8}$ &$5.34\,10^{8}$ \\ 
$\nu_{\element[][15]{O}}$    &$4.96\,10^{8}$ &$5.01\,10^{8}$ &$4.57\,10^{8}$ \\ 
$\nu_{\element[][17]{F}}$    &$6.41\,10^{6}$ &$3.15\,10^{6}$ &$5.74\,10^{6}$ \\ 
\\
$\Phi_{\rm Ga}$              & 130.1 & 128.4 & 129.8 
&$77.75\pm6.2^{+4.3}_{-4.7}$\\
$\Phi_{\rm Cl}$              &8.31   & 7.71  & 8.82  &$2.55\pm 0.25 $\\
$\Phi_{\rm Ka}$              &0.61   &0.55   &0.66   &$0.29\pm 0.02$\\
\\ \hline
\end{tabular}
\end{table}

\subsection{Neutrinos}
Table~\ref{tab:2} gives the predicted neutrino fluxes at earth level and the
expected fluxes for the three neutrino experiments namely,
chlorine (e.g. Davis \cite{da94}), gallium (e.g. Hampel et al. \cite{ha99})
and Kamiokande (e.g. Fukuda et al. \cite{fu96}), 
computed according to Berthomieu et al. (\cite{bpml93}). The gallium and
chlorine absorption cross sections have been taken respectively
from Bahcall (\cite{b97}) and Bahcall et al. (\cite{bla96}).
The $hep$ flux which may be important (e.g. Bahcall \& Krastev \cite{bk98},
Fiorentini et al. \cite{fb98})
in the neutrino spectrum measurements by
the SuperKamiokande, SNO and Icarus experiments is not listed.
With respect to C88, 
due to hotter core, $\nu_{\element[][7]{Be}}$ and CNO neutrino fluxes
are enhanced in A98 and N99 and, as expected, $\nu_{\rm pp}$ is slightly
reduced.

Despite larger temperatures in the core we obtained,
for the models A98 and N99 with respect to the model C88,
the expected decreases of the $\nu_{\element[][8]{B}}$
boron neutrino fluxes owing to their reduced rate
of the reaction $\element[][7]{Be}(p,\gamma)\element[][8]{B}^*$.
The introduction of N99 reaction rates relatively to
A98 induces an increase of $+10\%$ of $\nu_{\element[][8]{B}}$.
The effect is significant on the flux measured by the
chlorine and Kamiokande experiments.
Note that Table~\ref{tab:2} reveals that the neutrino fluxes at earth level for 
A98 are
very similar than those given Table 1 in Bahcall et al. (\cite{bbp98}).
They differ only by
few percent for $\nu_{\element[][17]{F}}$ owing to the large abundance of 
\element[][16]{O} resulting from the great efficiency of \element[][15]{N} burning
which about double the fraction of termination
of the NO part of the CNO bi-cycle. Despite the fact that the two stellar
evolution programs are entirely independent of each other,
when the same nuclear reaction rates are
used (A98), the most important neutrino fluxes 
$\nu_{\rm pp}$, $\nu_{\rm pep}$, $\nu_{\element[][7]{Be}}$,
$\nu_{\element[][8]{B}}$, $\nu_{\element[][13]{N}}$,
 and $\nu_{\element[][15]{O}}$ all agree to better than 1\%. 

\begin{figure}
\psfig{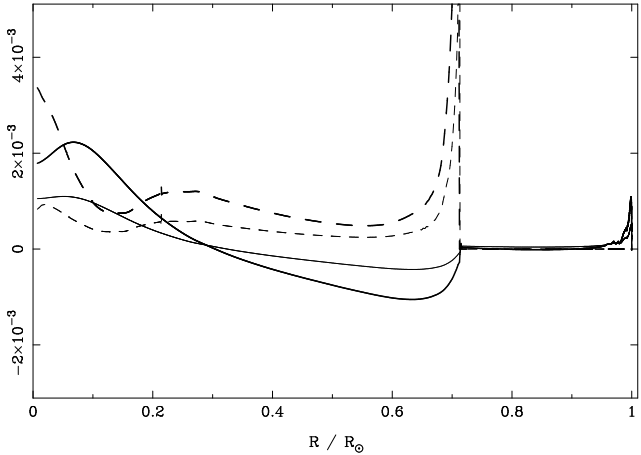}
\caption{
Relative frequency differences of sound speed (full)
and square of Brunt-V\"aiss\"al\"a frequency divided by 10 (dashed)
between N99 (thick) and C88, and A98 (thin) and C88.
}\label{fig:vais}
\end{figure}

\begin{figure}
\psfig{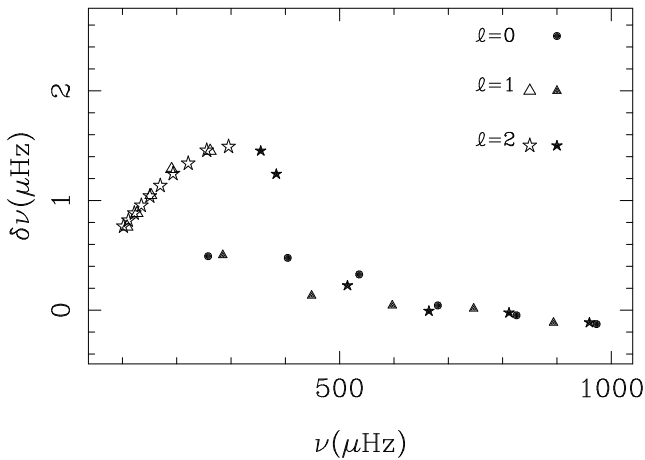}
\caption{
Frequency differences (in $\mu$Hz) in the low frequency range 
between N99 and C88 for modes of degree $\ell=0,\,1,\,2$.
Open symbols denote g-modes and, full symbols, f-modes and p-modes.
}\label{fig:pmod}
\end{figure}

\begin{figure}
\psfig{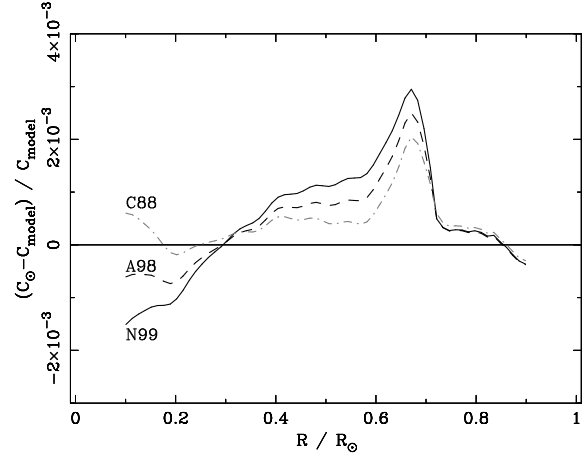}
\caption{
Relative difference in sound velocity between the Sun and the calibrated
models N99 (full), A98 (dashed) and C88 (dot-dash-dot-dash).
}\label{fig:vson}
\end{figure}

\subsection{Seismological comparison}
The seismic properties of the solar model are mainly related 
to the profile of sound-speed ($resp.$ Brunt-V\"aiss\"al\"a frequency)
as far as p-modes ($resp.$ g-modes) are concerned.
Figure \ref{fig:vais} shows that the models N99 and, to a lesser extend A98,
compared to C88, have a larger sound speed 
in the central core below 0.3 solar radius by $+0.2\%$ ($resp.$ $+0.1\%$),
and a
smaller one by $-0.1\%$ ($resp.$ $-0.05\%$) just below the convection zone. 
Table~\ref{tab:1} shows a small increase of small low degree differences
$\delta\nu_{02}$  
and $\delta\nu_{13}$, as defined Sec.~\ref{sec:phy} in relation with the difference of sound speed in the 
solar core. There the relative differences in the Brunt-V\"aiss\"al\"a frequency 
between models  N99 and C88 are larger by
a few  percents, i.e. one order of 
magnitude larger than the sound speed differences. The 
increase of the Brunt-V\"aiss\"al\"a frequency
leads to a $-1\%$ smaller value of $P_0$, the characteristic
spacing period of g-modes. The differences between A98 and C88
are three times smaller.
This change in Brunt-V\"aiss\"al\"a frequency influences
much the low frequency modes for frequency less than 1\,mHz,
i.e. the low radial order p-modes, the f- and g-modes. 
Consequently the frequency differences of the
low degree p-modes between models N99 and C88 
vary from $-0.1\,\mu$Hz to $-0.25\,\mu$Hz
when the frequency increases from  1\,mHz to 5\,mHz, with a minimum value of 
$-0.5\,\mu$Hz around 2 mHz.
The normalized frequencies differences for p-modes of degree 
$\ell=3$ to $\ell=150$ are negative and change by
less than $-1\,\mu$Hz in the observed range.
As expected, the change of nuclear reaction rates do not modify the frequency
of oscillation of degree larger than 70.

Figure~\ref{fig:pmod} shows the frequencies differences in low frequency range
between N99 and A98 and Table~\ref{tab:3} gives the frequencies of g-
and p-modes for $\ell=0$ to $\ell=2$ in the same frequency range.
In the low frequency range $400\,\mu$Hz -- 1\,mHz, it appears
that the p-mode frequencies
are changed by less than $0.5\,\mu$Hz between N99 and C88,
i.e. by less than $+0.1\%$, with an effect larger at lower frequency. 
Below $200\,\mu$Hz, the oscillations are gravity modes with an 
asymptotic behavior, and the relative period differences are almost 
proportional to $P_0$
given in Table~\ref{tab:3} (see Provost et al. \cite{pbm98} for details).
Between 200 to $400\,\mu$Hz, the oscillations are gravity modes, or f- and 
p1-modes and they are 
more influenced by the change of the Brunt-V\"aiss\"al\"a frequency in 
the solar core, induced 
by changes in nuclear reactions, except the p1-modes for $\ell=0$ and $\ell=1$. 
Frequencies shifts are much larger, 
of the order of  1 to $1.5\,\mu$Hz, when the frequency varies from $200\,\mu$Hz 
to $400\,\mu$Hz, i.e. about $1\%$, when comparing the models 
N99 and C88. 

In the range $0.1\,R_\odot\la R\la 0.9\,R_\odot$ where the inversions of
the helioseismic data are reliable,
the sound speed of the three models has been compared with the seismic
sound speed experimental results of Turck-Chi\`eze et al.
(\cite{t97}). Figure~\ref{fig:vson} shows that the
relative differences
are below a few $10^{-3}$. 
The discrepancy between the Sun and the models is larger for model N99 with
sound speed too small just below the convection zone and too large in the core.
 Table~\ref{tab:1} shows that
it is the same for the quantities $\delta\nu_{02}$ and 
$\delta\nu_{13}$ of the models compared to the corresponding 
observed values 
$\delta\nu_{n,\ell}$ derived from GOLF (Grec et al. \cite{gr97}) and
VIRGO/LOI (Fr\"ohlich et al. \cite{f97}) observations on SoHO.

\begin{table}
\caption[]{
Frequency of g-modes and p-modes in the range $100\,\mu$Hz--2\,mHz.
$\cal T$ is the
type of the mode labelled by the radial order.
}\label{tab:3}
\begin{tabular}{llllllllllll}  \hline \\
$\ell$& C88    &  N99   &$\cal T$&$\ell$& C88    &  N99   &$\cal T$\\ \\ \hline \\
0     &257.82  & 258.31 &p1  &2     &101.88  &102.64  &g10 \\
      & 404.32 & 404.79 &p2  &      &111.07  &111.88  &g9\\
      &535.98  & 536.30 &p3  &      &121.91  &122.80  &g8\\
      &680.67  & 680.71 &p4  &      &134.87  &135.82  &g7\\
      &825.56  & 825.52 &p5  &      &150.51  &151.55  &g6\\
      &972.95  & 972.82 &p6  &      &169.69  &170.83  &g5\\
      &1118.36 &1118.15 &p7  &      &193.24  & 194.49 &g4 \\
      &1263.78 &1263.60 &p8  &      &221.23  & 222.57 &g3\\
      &1407.86 &1407.58 &p9  &      &255.22  & 256.67 &g2\\
      &1548.78 &1548.51 &p10 &      &295.44  & 296.93 &g1\\
      &1687.12 &1686.81 &p11 &      &354.60  & 356.06 & f\\
      &1822.51 &1822.14 &p12 &      &383.36  &384.60  &p1 \\
      &1957.79 &1957.44 &p13 &      &514.35  & 514.58 &p2\\
      &        &        &    &      &664.33  & 664.33 &p3 \\      
1     & 108.55 & 109.31 &g5  &      &811.77  & 811.75 &p4  \\
      &127.04  & 127.93 &g4  &      &959.86  & 959.75 &p5  \\
      & 152.45 & 153.50 &g3  &      &1105.18 &1105.01 &p6  \\
      & 190.56 & 191.85 &g2  &      &1250.78 &1250.60 &p7  \\
      & 261.62 & 263.07 &g1  &      &1394.73 &1394.46 &p8  \\
      &284.77  & 285.27 &p1  &      &1536.06 &1535.80 &p9  \\
      &448.34  & 448.48 &p2  &      &1674.78 &1674.44 &p10  \\
      & 596.90 &596.95  &p3  &      &1810.37 &1810.01 &p11  \\
      &746.65  & 746.66 &p4  &      &1946.00 &1945.63 &p12  \\ 
      &893.71  & 893.60 &p5  \\      
      &1039.63 &1039.52 &p6  \\
      &1185.68 &1185.47 &p7  \\
      &1329.80 &1329.57 &p8  \\
      &1473.13 &1472.88 &p9  \\
      &1612.86 &1612.53 &p10 \\
      &1749.57 &1749.25 &p11 \\
      &1885.34 &1884.94 &p12 \\ 
\\ \hline
\end{tabular}
\end{table}

\section{Discussion and conclusions}\label{sec:dis}
We have compared the structure, the neutrino fluxes,
the chemical composition profiles
and the helioseismological properties of
calibrated standard solar models computed with the adopted
nuclear reaction rates of the European compilation NACRE
(Angulo et al. \cite{a99}) with those of calibrated solar models computed
with the nuclear reaction rates of Caughlan \& Fowler (\cite{cf88}) and 
Adelberger et al. (\cite{a98}).

Roughly speaking, the thermonuclear
reaction rates of PP chains adopted by NACRE and, but in less extend,
by Adelberger et al.,
are slightly less efficient than those adopted by Caughlan \& Fowler.
The calibration generates models with cores of larger temperature,
density, helium content and sound speed with the concomitant
increase of the neutrino fluxes, except for $\nu_{\rm pp}$ and
$\nu_{\element[][8]{B}}$; for this last one, the decrease is due to the smaller
rate of the reaction $\element[][7]{Be}(p,\gamma)\element[][8]{B^*}$. Thus
the predicted neutrino fluxes are reduced for the chlorine and Kamiokande
experiments, but almost unchanged for gallium. For Kamiokande and 
chlorine, N99 predicts intermediate values between A98 and C88.

The introduction of the NACRE thermonuclear
rates increases the discrepancy between
predicted and observed sound velocity profiles
between the Sun and the
models, both below the convection zone and in the solar core.
These relative differences, though at the level of a
few thousandths, are smaller for the model
computed with the reaction rates of
Caughlan \& Fowler, the increase is $\sim+0.5\%$ for C98 and $\sim+1\%$ for N99.
The radius at the base of the solar
convection zone is in good agreement with the observed value
for all models.

Though NACRE adopts an enhanced rate for the reaction of lithium burning 
$\element[][7]{Li}(p,\alpha)\element[][4]{He}$, the predicted depletion
of photospheric lithium remains too small to fit the observed value.

The differences between
calibrated solar models computed with the adopted thermonuclear
reaction rates of the two new compilations are rather small.  It is
not really possible to make any choice between them. 
From an increase of the accuracy of the observed p-mode frequencies and from
a hopefully detection of g-modes one can expect to improve our knowledge
on the stratification of the solar core in the goal
to validate, in the low energy regime,
the thermonuclear reaction rates and their concomitant neutrino generation.

Thanks are due to the NACRE's work which also provides these estimates of
uncertainties on the adopted rates. These new features, now available to
the user, are important to constraint the solar
model. We are investigating this point in a work in progress.

\begin{acknowledgements}
It is a pleasure to thank the referee Pr. J.N. Bahcall for bringing several
references to our attention, helping us to clarify several points
and making several constructive
suggestions which have improved the paper.
This work has been performed using the computing facilities 
provided by the OCA program
``Simulations Interactives et Visualisation en Astronomie et M\'ecanique 
(SIVAM)''. W. D\"appen is acknowledged for kindly providing the MHD
package of equation of state.
\end{acknowledgements}

\end{document}